\documentclass[aps,prb]{revtex4}
\usepackage{graphicx}
\newcommand{\micron}[1]{\mbox{#1~$\mu$m}}

\begin{document}

\title{Low density fragile states in cohesive powders}
\author{Paul B. Umbanhowar}
\email[]{umbanhowar@northwestern.edu}
\affiliation{Department of Physics and Astronomy, Northwestern University, Evanston IL, 60208, USA.}
\author{Daniel I. Goldman}
\email[]{digoldma@berkeley.edu}
\affiliation{Department of Integrative Biology,
University of California at Berkeley, Berkeley CA, 94720, USA.}

\date{\today}

\begin{abstract}
We discuss the difference between cohesive and non-cohesive granular
media in the context of a recent report of ``dry quicksand." Weak
low density states with properties like dry quicksand are readily
formed in common household powders. In contrast, such states cannot
be formed in cohesionless granular media such as ordinary sand.
\end{abstract}

\maketitle

Sand and other granular media have the intriguing ability to exhibit
properties of both fluids and solids --- readily poured, yet able to
hold a shape and bear weight. Recently, a scientific report of ``dry
quicksand", capable of swallowing desert travelers\cite{lohArau}
captured the public's imagination and was widely discussed in
popular news media. Such broad interest is not surprising given that
ordinary sand is capable of supporting even heavy loads. However,
far from being exotic and associated only with remote desert regions
and carefully prepared experiments, dry quicksand's distinguishing
traits are characteristic of ordinary cohesive powders, which can be
found in the home, on ski slopes, and even on
asteroids\cite{thomaspc} and the moon\cite{wilhelms}.
\bigskip

In the work mentioned above, Lohse {\em et~al.}\cite{lohArau}
describe experiments in which a 2~cm diameter, 133~g ball is
released at the surface of a bed of $40~\mu$m diameter quartz
grains. Without special preparation the bed supports the ball.
However, after forcing air upward through the bed, which reduces the
solid volume fraction $\phi$ to $41\%$, the ball sinks to a depth of
about five diameters producing a jet of sand that shoots into the
air\cite{lohAber,thoAshe}. The authors call this fragile state ``dry
quicksand."
\bigskip

We were able to produce nearly identical behavior without elaborate
preparation using common powders such as confectioners sugar and
cake flour. We found that corn starch based Johnson's Baby Powder
produced the largest jets, as shown in Fig.~1. A steel ball released
at the surface of the powder ($\phi \approx 39\%$) fell to the
bottom of the container, and a well defined jet emerged. In a 30~cm
deep powder of hollow \micron{5-200} diameter glass beads, the
material was so fragile that the ball bounced repeatedly on the
bottom of the container. The weak low volume fraction states
required for jet formation were easily prepared by gently tumbling
the powder or by shaking the powder together with the ball. For all
the materials we tested tapping the container on a solid surface
compacted the powder which prevented the ball from sinking.
\bigskip

In {\em non-cohesive}, disordered particulate media, such as
ordinary sand, the low volume fraction states described above are
unattainable. For an idealized granular material composed of
identical spheres the volume fraction of disordered configurations
is bounded by two important limits: the maximum volume fraction
state called Random Close Packed with $\phi_{\mathrm{rcp}}\approx
64\%$, and the minimum volume fraction state called Random Loose
Packed with $\phi_{\mathrm{rlp}}= 55 \pm 0.5 \%$.  The latter state
can be realized by allowing particles to settle in a nearly
density-matched fluid\cite{onoAlin}. In contrast, aeration of
spherical glass beads (as small as \micron{50} diameter) yields a
larger minimum volume fraction of $\phi = 59 \pm 0.4\%$ independent
of particle size\cite{ojhAmen,golAswi05}. It is also important to
note that random configurations of non-spherical ellipsoidal
particles such as M\&M candies have significantly higher values of
$\phi_{\mathrm{rcp}}$\cite{chaikin} and are expected to have
correspondingly larger values of $\phi_{\mathrm{rlp}}$ as
well\cite{weitz}.
\bigskip

In {\em cohesive} media, however, fragile, loosely packed states
with $\phi < \phi_{\mathrm{rlp}}$ are common when the attractive
forces between grains ({\it e.g.} van der Waals, electrostatic, and
capillary due to the presence of interstitial fluid) exceed the
grain weight\cite{rietemabook}. There are many reports (see, for
example, Valverde {\em et~al.} and references therein\cite{valAcas})
of such low volume fraction cohesive powders. These authors describe
measurements of the tensile strength of beds of \mbox{9 $\mu$m}
diameter toner particle with $\phi$ less than $35\%$ achieved by
aeration --- in non-cohesive granular materials the tensile strength
is strictly zero. Ballistic deposition can create states with $\phi$
as low as $15\%$~\cite{bluAsch}. The substantial variation in fine
powder density as a function of its preparation history is
characterized by its own number: the Hausner ratio measures the
ratio of aerated to tapped powder density\cite{hausner}.
\bigskip

For fixed attractive mechanisms and fixed material density of the
particles, the transition between a non-cohesive and a cohesive
material depends primarily on the particle size --- cohesive powders
typically have particle diameters less than \mbox{10 $\mu$m},
whereas freely flowing, non-cohesive granular materials are the norm
for diameters greater than \mbox{100 $\mu$m}. The low $\phi \approx
41\%$ measured by Lohse {\em et al.} strongly suggests that the
quartz grains used in their study were just small enough to form a
cohesive powder. Further, low $\phi$ states are widely known to be
weak whether they are composed of snowflakes\cite{snow} or metal
particles\cite{metals}---few people would be surprised to see a
measuring spoon vanish into sieved flour. In addition, the
possibility of the Apollo astronauts being swallowed by loosely
packed cohesive moon dust was considered a genuine risk for the
first lunar landing~\cite{wilhelms}; here on earth, it is common to
sink deeply into dry powdered snow.
\bigskip

Studies like those of Lohse {\em et~al.}~\cite{lohArau} have
introduced the fascinating behaviors of powders to a wide audience,
and they also demonstrate the still largely unappreciated and
unexplored physics of these common materials. There is much to learn
from and about cohesive powders and the non-cohesive to cohesive
transition, and many simple, basic, and low cost investigations are
waiting to be done. For example, how does the volume fraction of the
loosest possible stable state depend on the particle shape, size,
and size distribution? In what way do external perturbations such as
vibration or variation of temperature affect volume fraction and
strength? Can cohesive forces be tuned by controlling electrostatic
interactions\cite{gryz}, varying the amount of interstitial
fluid\cite{kudrolli}, or by creating engineered particles with
adhesive hairs like those on the foot of a gecko\cite{autAlia}?
\bigskip

In summary, ensembles of cohesionless particles like those found in
sand dunes and sandboxes have been the subject of much recent
attention and exhibit many interesting behaviors\cite{njbrmp};
however, such materials do not form the fragile states necessary to
create dry quicksand or other states with $\phi <
\phi_{\mathrm{rlp}}$ where attractive forces play a leading role.

\begin{figure}
\begin{center}
\includegraphics[width=3in]{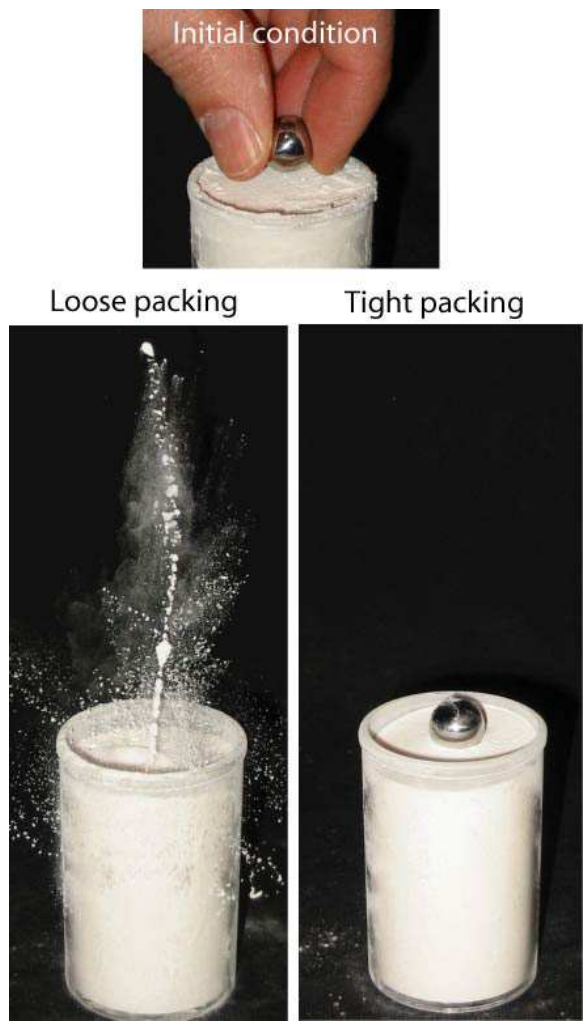}
\caption{\label{figure1} A 1.27 cm diameter steel ball released (top
panel) at the surface of vial filled with baby powder (maximum
particle diameter \mbox{45 $\mu$m}) filled vial rapidly penetrates
to the bottom and a jet forms (left panel), demonstrating that the
effects obtained in ``dry quicksand"~\cite{lohArau} are common in
loosely packed cohesive powders. When the powder is packed by
tapping the container a few times on a solid surface, it supports
the weight of the ball (right panel). The density ratio of the loose
to the tapped state (the Hausner ratio) is 0.8; the solid volume
fraction of the loose state is approximately $39\%$. Images (left
and right panels) are taken about 150~msec after the release of the
ball.}
\end{center}
\end{figure}

\bibliography{Powderbib}
\end{document}